\documentclass[12pt]{iopart}

\newcommand{\eqn}[1]{(\ref{#1})}

\renewcommand{\d}{\delta}
\newcommand{\g}{\gamma}

\newcommand{\m}{\mu}

\newcommand{\n}{\nu}

\newcommand{\ft}[2]{{\textstyle\frac{#1}{#2}}}
\newcommand{\be}{\begin{equation}}
\newcommand{\eq}{\end{equation}}

\newcommand{\pa}{\partial}

\def\beqa{\begin{eqnarray}}
\def\eeqa{\end{eqnarray}}

\begin{document}
\jl{6}

\begin{center}
\hfill AEI-1999-27\\
\hfill THU-99/29\\[1mm]
\hfill {\tt hep-th/9910179}\\
\end{center}

\title{Area law corrections from state counting and supergravity }

\author{Gabriel Lopes Cardoso\dag, Bernard de Wit\ddag\footnote[5]{Talk
presented by B. de Wit at Strings '99, Potsdam, July 19-24 1999}
and
Thomas Mohaupt\S}

\address{\dag\ddag 
Institute for Theoretical Physics, Utrecht University,
3508 TA Utrecht, The Netherlands}
\address{\ddag
Max-Planck-Institut f\"ur Gravitationsphysik,
Albert-Einstein-Institut,
Am M\"uhlenberg 1, D-14476 Golm, Germany}
\address{\S 
Martin-Luther-Universit\"at Halle-Wittenberg, 
Fachbereich Physik,
D-06099 Halle, Germany}    

\begin{abstract}
Modifications of the Bekenstein-Hawking area law for black holes 
are crucial in order to find agreement between the microscopic 
entropy based on state counting and the macroscopic entropy based
on an effective field theory computation.  We discuss this and related
issues for  the case of four-dimensional $N=2$ super\-symmetric black
holes. We also briefly comment on the state counting for $N=4$ and
$N=8$ black holes. 
\end{abstract}


\submitted

\maketitle

\section{Introduction}
A microscopic derivation of the entropy of certain extremal black holes 
has recently become available in the context of string theory 
\cite{StromVafa}-\cite{Vaf}.  For 
four-dimensional extremal black holes in the limit
of large electric/magnetic charges $Q$, 
the microscopic entropy is generically of the form
\beqa
{\cal S}_{\rm micro} \sim \sqrt{Q^4} \;\;.
\label{entromicro}
\eeqa
This result agrees with the one obtained from macroscopic calculations
based on the corresponding effective field theories.  Here one first
constructs the associated black hole solution and then one computes
the macroscopic entropy according to the Bekenstein-Hawking area law 
\cite{Haw,Bek}.

In the context of string theory and M-theory, 
the microscopic entropy (\ref{entromicro})
is calculated by counting excitations of D-branes and M-branes.  In the context
of type-IIA compactifications on Calabi-Yau threefolds $CY_3$ 
extremal black holes
are microscopically represented by wrapping a D4-brane on a smooth
holomorphic four-cycle $P$ of the Calabi-Yau threefold and by
considering its
bound state with $|q_0|$ D0-branes.
In M-theory compactifications on $CY_3 \times S^1$, on the other hand,
they are represented by 
five-branes wrapped on $P \times S^1$, with $|q_0|$ quanta of lightlike
momentum along the circle $S^1$.  The massless excitations of the five-brane
wrapped on $P$ are described by a $(0,4)$ two-dimensional conformal theory
\cite{MalStrWit}, and the degeneracy of states of this conformal field
theory yields the microscopic entropy according to Cardy's formula,
\beqa
d(|q_{0}|,c_{L}) \approx \exp({\cal S}_{\rm {micro}}) \approx 
\exp\Big( 2 \pi \sqrt{ \ft16{|q_{0}| c_{L}} } \,\Big)\;.
\label{cardmicro}
\eeqa
Here $|q_{0}|$ is taken to be large 
and $c_{L}$ denotes the central charge for the left-moving sector.
Evaluation of (\ref{cardmicro}) for the case of a five-brane
wrapped around $P \times S^1$ yields \cite{MalStrWit}, 
\beqa
{\cal S}_{\rm {micro}} = 2 \pi \sqrt{ \ft16{|q_{0}|} \Big(
C_{ABC}\, p^{A} p^{B} p^C + c_{2A}\, p^A \Big) } \;,
\label{genericentropia} 
\eeqa
where $C_{ABC}$ and $c_{2A}$
denote the triple intersection numbers and the second Chern class numbers
of the Calabi-Yau threefold, respectively.  The charges $p^A$ denote the
expansion coefficients of the four-cycle $P$ in a homology basis $\Sigma_A$
of four-cycles, $P = p^A \Sigma_A$.  In obtaining (\ref{genericentropia})
the topological data of the four-cycle $P$ have been expressed in terms
of topological data of the Calabi-Yau threefold.

The Calabi-Yau compactification reduces the number of supersymmetries
to eight, 
so that the black hole solution has $N=2$ supersymmetry at spatial
infinity. One may also consider compactifications on $K3\times T^2$
or on $T^6$ with the five-brane wrapping a four-cycle. In that case one
has $N=4$ or $N=8$ supersymmetry at spatial infinity while the
massless excitations are still described by a $(0,4)$ two-dimensional
conformal field theory. Hence one obtains entropy formulae similar to 
\eqn{genericentropia}. We will return to them later. 

Inspection of (\ref{genericentropia}) shows that, for large charges $p^A$,  
there are subleading
corrections (proportional to $c_{2A}$) 
to the microscopic entropy.
It was argued in \cite{MalStrWit,Vaf} that these deviations in
the entropy formula should, at the macroscopic level, arise 
from terms in the effective action proportional to the square of the 
Weyl tensor, with coefficients linearly related to the second Chern class
of the Calabi-Yau threefold.

The associated four-dimensional effective field theory is based on $N=2$
supergravity coupled to a number of vector multiplets whose gauge
fields are 
associated with electric and magnetic charges, denoted by
$q_I$ and $p^I$, respectively. The theory incorporates, in a
systematic fashion, the phenomenon of electric/magnetic duality,
according to which the electric and magnetic charges can be
interchanged and/or rotated and it includes higher-derivative
couplings with among them a certain class of 
terms quadratic in the Weyl tensor.  The effective theory is thus complicated
and depends on many fields.
How can the macroscopic
description of extremal $N=2$ 
black holes then be so constrained and systematic as
to precisely reproduce the results from the counting of microstates such as
(\ref{genericentropia})?  The crucial ingredient that is responsible
for the remarkable restrictions on the entropy formulae obtained on the
basis of these complicated effective field theories is the enhancement
to full supersymmetry at the horizon.  The black hole 
solutions that we consider are static, 
rotationally symmetric solitonic interpolations between 
two $N=2$ supersymmetric groundstates: flat Minkowski spacetime at spatial 
infinity and Bertotti-Robinson spacetime at the horizon 
\cite{Gib,CarDeWMoh}. The 
interpolating solution preserves $N=1$ supersymmetry so that we 
are dealing with a BPS configuration and the black hole is extremal. 
The interpolating solution  depends, generically, on the electric and
magnetic charges as well as on the values of the moduli fields at 
spatial infinity. The supersymmetry enhancement at the horizon is
responsible for the    
fixed-point behaviour of the moduli forcing them to take 
certain values depending on the electric/magnetic charges at the horizon
\cite{FerKalStr,Moore}. The precise relation can be deduced from
electric-magnetic duality considerations 
\cite{BCDWKLM,CarDeWMoh}.  
The near-horizon geometry is thus entirely determined
in terms of the charges carried by the black hole, and so is the entropy.

\section{Supersymmetric black hole solutions}

The supergravity Lagrangians that give rise to
these extremal black hole solutions are based on the coupling of $n$
vector multiplets to $N=2$ 
supergravity. They contain various other couplings, such as
those associated with hypermultiplets, which play only a
limited role in the following and will be omitted.
The construction of the coupling of vector multiplets to $N=2$
supergravity utilizes the so-called superconformal multiplet calculus
\cite{DWVHVPL} which enables one to straightforwardly include the 
interactions proportional to the square of the Weyl tensor. Let us
recall that the covariant fields of a vector 
multiplet, the field strength of the vector gauge field, 
a complex scalar, a doublet of gaugini and a triplet of auxiliary
scalar fields, constitute a restricted chiral multiplet. The complex
scalar fields are denoted by $X^I$. We consider only abelian vector
multiplets, which we label 
by $I= 0,1,\ldots,n$. The extra vector multiplet is required to provide the
graviphoton field of supergravity. The supersymmetric (Wilsonian)
action is encoded 
in a holomorphic function $F(X)$ of the scalars (or, in superspace, of
the corresponding chiral superfields). Under electric/magnetic duality
transformations the function $F(X)$ changes, but the corresponding
equations of motion and Bianchi identities remain the same. 

In the superconformal framework there is another multiplet, the so-called
Weyl multiplet, which comprises the gravitational degrees of freedom, namely
the graviton, two gravitini as well as various other superconformal
gauge fields and also some auxiliary fields.  One of these
auxiliary fields is an anti-selfdual Lorentz tensor field $T^{ab\, i j}$,
where $i,j=1,2$ denote chiral $SU(2)$ indices, which occurs in the
gravitino transformation law according to $\d\psi^i_\m = 2{\cal
D}_\m\epsilon^i -{1\over8} \g_{ab}\g_\m \,\epsilon_j\,T^{abij} + \cdots$. 
The covariant quantities of the Weyl multiplet also reside in a
reduced chiral multiplet, denoted 
by $W^{ab\, i j }$, from which one constructs the unreduced
chiral multiplet $W^2 = (W^{ab\,i j } \varepsilon_{ij})^2$ \cite{BDRDW}.
The lowest component field of $W^2$ is equal to ${\hat A} = 
(T^{ab\, i j } \varepsilon_{ij})^2$.
Because $W^2$ is also a chiral multiplet, we can simply include
interactions between the vector multiplets and the Weyl multiplet by
extending the holomorphic function to a function that depends both on
$X^I$ and $\hat A$. However, this function must be homogeneous of degree
two and thus satisfies $X^I F_I + 2 {\hat A}\, F_{\hat A} = 2 F$, where
$F_I = \partial F/\pa {X^I}$, $F_{\hat A} = \partial F/\pa {\hat
A}$. The most prominent interaction term that is induced by the $\hat
A$-dependence is quadratic in the Weyl tensor and proportional to the
derivative $F_{\hat A}$. Note that 
there are no terms proportional to the derivative of the Riemann
tensor. 

Our first task is to find all $N=2$ supersymmetric field configurations
(in the full off-shell theory) that are consistent with the  static,
spherically symmetric, black hole geometry, which in isotropic coordinates
$(t,r,\phi,\theta)$ is described  by the line element 
\be
ds^2 = - \e^{2g(r)}dt^2 + \e^{2 f(r)} ( dr^2 + r^2  d \Omega^2 )\,.
\label{StaticMetric}
\eq
An off-shell analysis which includes general interactions with the
Weyl multiplet \cite{CarDeWMoh} reveals that there exists 
only one class of fully supersymmetric solutions, namely the 
Bertotti-Robinson spacetime corresponding to $adS_2\times S^2$. This
geometry is thus relevant for the black hole 
near the horizon or at spatial infinity (where the anti-de Sitter
radius tends to infinity so that one is dealing with flat Minkowski
spacetime). It is important to stress here that we do not explicitly
use the action or field equations (except for the abelian gauge
fields which are induced by the presence of the electric/magnetic
black hole charges), as everything is encoded in the 
function $F(X,\hat A)$. The use of an off-shell formulation is
essential in view of the fact that the action is extremely 
complicated and generates an infinite sequence of higher-derivative
interactions upon integrating out the auxiliary fields. In this way 
one obtains, for instance, an infinite series of terms proportional to
the square of the Weyl tensor times powers of the field strengths
associated with the vector multiplets. In view of the maximal 
supersymmetry the corresponding field equations must be satisfied. 
 
The analysis of \cite{CarDeWMoh} shows 
that the $X^I$ and ${\hat A}$ must be constant (that is, in a certain
gauge; in principle only appropriate ratios are determined in view of the
invariance under local dilatations of the superconformal formulation)
and that $\e^{2g(r)} =
\e^{-2f(r)} = \e^{-{\cal K}} \, \vert Z\vert^{-2} \,r^2$, where 
\beqa
Z &=&  \e^{ {\cal K}/2} \, (p^I F_I (X,{\hat A}) - q_I X^I) \,,
\nonumber \\
\e^{-\cal K} &=& i [\bar{X}^I F_I (X,{\hat A}) -  
\bar{F}_I ({\bar X}, {\bar {\hat A}}) X^I ] \,.
\label{z}
\eeqa
This shows that we are dealing with a spacetime geometry that is of
the Bertotti-Robinson type. 
Note the dependence on the black hole magnetic and electric charges
$(p^I, q_I)$.  
The quantities $Z$ and $\cal K$ are both generically 
non-vanishing and constant. It was also found that 
$T^{01\,ij} = - i\, T^{23\,ij} = 2\,
\varepsilon^{ij} \, \e^{-{\cal K}/2}\, \bar Z^{-1}$,
while all other
components of $T^{ab\,ij}$ 
vanish. Therefore we have $\hat A =-  64\,\e^{-{\cal K}}\,\bar Z^{-2}$. 

Hence the fully supersymmetric field configurations are characterized
in terms of the (constant) moduli $X^I$ (or rather, their ratios) and
the electric/magnetic 
charges. However, when the field configuration satisfies the field
equations and the Bianchi identities, which we know must be the case,
then it must also be consistent 
with electric/magnetic duality. These equivalence transformations take
the form of symplectic 
${\rm SP}(2n+2;{\rm \bf Z})$ transformations.  Now we observe that
both $(p^I,q_J)$ and $(X^I,F_J)$ transform as symplectic vectors under
duality transformations \cite{deWit1}. Since they are the only such
vectors left in these supersymmetric configurations, they must satisfy
a proportionality relation, which in 
principle determines the $X^I$ in terms of the charges
\cite{BCDWKLM,CarDeWMoh}. Therefore fully supersymmetric field
configurations are completely parametrized in terms of the
charges.  Observe 
that this observation already indicates that, also in the presence of
higher-derivative interactions, the moduli will exhibit fixed-point
behaviour at the horizon. An explicit proof of this will be presented
elsewhere \cite{CEWKM}. 

What remains is to calculate the entropy for  particular black hole
solutions which interpolate between the two different fully
supersymmetric field configurations at spatial infinity and at the
horizon. Since the behaviour at the horizon is completely determined in
terms of the charges, the resulting entropy formula will only depend
on these charges. However, if one computes the 
macroscopic entropy  for a black hole of the type considered in
\cite{MalStrWit,Vaf} by using the area law of Bekenstein and
Hawking, then one discovers \cite{BCDWLMS} that the resulting expression 
does {\it not} agree with the expression for the microscopic entropy
(\ref{genericentropia}).  Thus, in order to
obtain agreement with the counting of microstates provided by string
theory, one is forced to depart from the area law.  For that reason we
adopt Wald's proposal for the entropy which ensures the validity of the
first law of black hole mechanics for more generic field
theories. This proposal is based on the existence of a Noether charge 
associated with an isometry evaluated at the corresponding  
Killing horizon \cite{Wald}. When evaluating this current subject to
the field equations, current conservation becomes trivial and the
current can be written as the divergence of an antisymmetric
tensor. This tensor, sometimes called the Noether potential, is a
local function of the fields and of the (arbitrary) gauge
transformation parameters. Its integral over the horizon yields the
macroscopic entropy. 

\section{Entropy as a Noether charge}

In order to elucidate Wald's Noether charge proposal let us first
briefly consider a simple three-dimensional
abelian gauge theory, with a gauge-invariant Lagrangian 
depending on the field strength $F_{\m\n}$,  its derivatives $\pa
_\rho F_{\m \n}$,  as well as on matter  
fields $\psi$ and first derivatives thereof. Furthermore we add a
Chern-Simons term (which acts as a topological mass term), so that the
total Lagrangian takes the form
\beqa
{\cal L}^{\rm total} = {\cal L}^{\rm inv}(F_{\m\n},\pa_\rho F_{\m\n},
 \psi, \nabla_{\!\mu\,}\psi) + c\,\varepsilon^{\m\n\rho} A_\m\,\pa_\n
 A_\rho \,,
\eeqa
where $\nabla_{\!\mu\,}\psi$ is the covariant derivative of
$\psi$ and $c$ is some constant. This Lagrangian is not gauge invariant but
changes into a total derivative, 
\beqa
\d_\xi{\cal L}^{\rm total} = \pa_\m N^\m(\phi,\xi) =
 c\,\varepsilon^{\m\n\rho}\,\pa_\m \xi \,\pa_\n A_\rho \,,
\eeqa
where generically $\phi$ denotes all the fields and $\xi$ denotes the
transformation parameter. 
For field configurations that satisfy the equations of motion, the
corresponding Noether current can be written in terms of a so-called
Noether potential ${\cal Q}^{\m\n}$, which in the case at hand reads 
\beqa
{\cal Q}^{\m\n}(\phi, \xi) = 2 \,{\cal L}^{\mu\nu}\, \xi 
- 2\,\pa_{\rho}{\cal
L}^{\rho, \mu\nu} \, \xi +  {\cal L}^{\rho,
\mu\nu}\, \pa_{\rho} \xi + 2c \,\varepsilon^{\m\n\rho}
A_\rho\,\xi  \,. \label{Npotential}
\eeqa
Here ${\cal L}^{\mu\nu}$ and ${\cal L}^{\rho, \mu\nu}$ denote the
derivatives of the action with respect to 
$F_{\mu\nu}$ and $\pa_{\rho} F_{\mu\nu}$, respectively. Observe that
the Bianchi identity implies ${\cal L}^{[\rho,\mu\nu]}=0$.  The
Noether potential, whose definition is not 
unambiguous,  is a local function of the fields and of the transformation
parameter $\xi$. Observe that ${\cal Q}_{\m\n}$ does not have to
vanish for field 
configurations that are invariant (in the case at hand, this would
imply $\pa_\m\xi = \xi\psi=0$). 
Modulo equations of motion the corresponding Noether current equals 
\beqa
J^\mu(\phi,\d_\xi\phi) = \pa_\n \,{\cal Q}^{\m\n}(\phi,\xi)\,.
\eeqa
Note that the current depends only on the gauge parameter through the
gauge variations $\d_\xi\phi$. This property is not automatic and in
order to realize this we made a particular choice for the Noether
potential, exploiting the ambiguity in its definition \cite{deWit2}

Integration of the Noether potential \eqn{Npotential} over the
boundary of some (spacelike) hypersurface leads to a surface charge,
which, when restricting the gauge transformation parameters to those
that leave the background invariant, is equal to the Noether charge in
the usual sense. In the case at hand this surface charge remains
constant under variations that continuously connect 
solutions of the equations of motion. Here we consider a continuous
variety of solutions of the field equations which are left invariant
under a corresponding variety of residual gauge
transformations. Hence, the parameters $\xi$ that characterize the
residual symmetry may change continuously with the solution. Denoting
the combined change of the solution $\phi$ and of the symmetry 
parameters $\xi$ by the variation $\hat \d$, one may thus write
\beqa
\hat\d (\d_\xi\phi) = 0\,.
\eeqa
In our example the Noether current can be written as a function of
$\phi$ and $\d_\xi\phi$, so that one knows that $\hat\d
J(\phi,\d_\xi\phi)$ remains proportional to $\d_\xi\phi$ and must
therefore vanish for the symmetric configurations. Consequently 
$\hat\d{\cal Q}^{\m\n} (\phi,\xi)$ must vanish up to a closed form,
$\pa_\rho \omega^{[\m\n\rho]}$, so that the surface charge obtained by
integration over a Cauchy surface $C$ with volume element ${\rm
d}\Omega^\m$, 
\beqa
\int_C \, {\rm d}\Omega_\m\,J^\m(\phi,\d_\xi\phi) = \oint_{\pa C}
\,{\rm d}\Sigma_{\m\n} \,{\cal Q}^{\m\n}(\phi,\xi) \,,
\eeqa
is constant under the variations induced by $\hat\d$. Observe that
the integrand on the right-hand side is in principle nonvanishing and
nonconstant, so that the constancy of the total surface charge
represents a nontrivial result. 

In general relativity one follows the same approach as in the above
example. The gauge transformations then take the form of 
diffeomorphisms and the residual gauge symmetries are associated with
Killing vectors. The Lagrangian is not invariant but transforms as a
density, which implies that $N^\m(\phi,\xi) \propto
\xi^\m\,{\cal L}$. Proceeding as in the  
example discussed above, the associated Noether current gives rise to 
a Noether potential. However, in this case there are a number of
complications when considering variations of the surface
charge. Another essential ingredient is that the boundary decomposes
into two disconnected parts for black hole solutions, one associated with
spatial infinity and one with the horizon. After identifying a surface 
charge that is constant under variations within a continuous variety
of solutions of the equations of motion, the
contributions coming from variations at spatial infinity must 
cancel against those 
coming from the horizon. It is this phenomenon that ensures the
validity of the first law of black hole mechanics: the contributions
originating from spatial infinity are related to variations of the
black hole mass and angular momentum, while the contributions
originating from the horizon are identified with the change of the
black hole entropy (see \cite{JacKaMy,IWald,WaldRev} for a review). In
this way one establishes a formula for the black hole entropy in terms
of the surface charge of the Noether potential over the horizon. 
When the Lagrangian depends arbitrarily on the Riemann tensor
$R_{\m\n\rho\sigma}$ (but not on its derivatives) and on matter fields
and their first-order derivatives, one can show that 
the entropy of a static black hole is given by \cite{Visser,JacKaMy,IWald}
\beqa
{\cal S}_{\rm macro} = \ft1{16}  \oint_{S^2} \;
\varepsilon_{ab}\,\varepsilon_{cd} \,  {\pa(8\pi {\cal L})\over
\pa R_{abcd}} \,,
\label{graventro}
\eeqa
where the epsilon tensors act in the subspace orthogonal to the
horizon associated with the time
and the radial coordinates; the factor $8\pi$ is related to our
normalization conventions for the Lagrangian.

With these results one can compute the macroscopic entropy of 
static, spherically symmetric,  $N=2$ supersymmetric black hole
solutions in the presence of higher-derivative interactions. In
view of the homogeneity of the function $F$, it is convenient to
introduce rescaled variables $Y^I =  \e^{{\cal K}/2}
\bar{Z} X^I$ and $\Upsilon = e^{ {\cal K} } 
\bar{Z}^2 \hat{A}$. At the horizon we must have $\Upsilon =
-64$. It then follows that the relation between the $Y^I$ and
the electric/magnetic charges of the black hole is given by 
$Y^I - \bar{Y}^I = ip^I$ and $F_I(Y,\Upsilon) - \bar{F}_I(\bar{Y}, 
\bar{\Upsilon}) = i q_I$.  On the other hand, it follows 
from (\ref{z}) that $|Z|^2 =
p^I F_I(Y,\Upsilon) - q_I Y^I$, which determines the value of $|Z|$
in terms of $(p^I, q_I)$. Subsequently one establishes that the
expression for the entropy takes the remarkably concise form
\cite{CarDeWMoh}, 
\be
{\cal S}_{\rm macro} 
= \pi \left[ |Z|^2 +4\, \mbox{Im} \,\Big(
\Upsilon\,F_{\Upsilon}(Y,\Upsilon)\Big) \right] 
\,,\quad 
\mbox{   where  } \Upsilon = - 64 \,.
\label{entropia}
\eq
In this formula the first term originates from the Bekenstein-Hawking
entropy contribution associated with the area, whereas
the second term is due to Wald's modification induced by the presence
of higher-derivative terms. Here we 
point out that this modification does not actually originate from the 
terms quadratic in the Weyl tensor, because the Weyl tensor vanishes
at the horizon, but from a term in the Lagrangian  proportional to the  
product of the Ricci tensor with the tensor field 
$T^{ab\,ij} T_{cd \,kl}$. Note that when switching on
higher-derivative interactions the value of 
$|Z|$ changes and hence also the horizon area changes.  There are thus 
two ways in which the presence of higher-derivative interactions
modifies the black hole entropy, 
namely by a change of the near-horizon geometry and by an explicit
deviation from the Bekenstein-Hawking area law.
Also note that the entropy (\ref{entropia}) is entirely determined
in terms of the charges carried by the black hole, ${\cal S} = {\cal
S} (q,p)$. Because of the homogeneity property of the function
$F(Y,\Upsilon)$ one can show that the macroscopic entropy
(\ref{entropia}) must be an {\it even} function of the charges.

\section{An $N=2$ example}

Let us then determine the macroscopic entropy of black hole solutions
arising in type-IIA string theory compactified on a Calabi-Yau
threefold, in the limit where the volume of the Calabi-Yau threefold
is taken to be large, and let us compare it with the result for the 
microscopic entropy (\ref{genericentropia}) obtained via state counting.
The associated homogeneous function
$F(Y,\Upsilon)$ is given by (with $I = 0, \ldots, n$ and  $A = 1,
\ldots, n$) 
\be
F(Y,\Upsilon) 
=- \frac{C_{ABC} \,Y^A Y^B Y^C}{6\, Y^0} - \frac{1}{24} \, \frac{1}{64}\;
c_{2A} \, \frac{Y^A}{Y^0} \; \Upsilon \;. 
\label{prep2a}
\eq
The Lagrangian associated with 
this homogeneous function contains a term proportional to 
the square of the Weyl tensor with coefficient $c_{2A} \, {\rm Im }\,
z^A$, where $z^A = {Y^A}/{Y^0}$.  Consider, in particular, black holes
carrying charges $q_0$ and $p^A$, only.  Solving the 
associated stabilization equations for $Y^I=Y^I(q,p)$ and substituting
the result into (\ref{entropia}) yields \cite{CarDeWMoh}
\beqa
{\cal S}_{\rm macro} = 
2 \pi \sqrt{\ft16 \, |q_0|  (C_{ABC} \,p^A p^B p^C + 
{c}_{2A} \, p^A) }\,,
\label{entrot2}
\eeqa
in exact agreement with 
the microscopic entropy formula (\ref{genericentropia}).
Thus, we see that the entropy obtained via state counting
\cite{MalStrWit,Vaf} is in accord with Wald's proposal for the
macroscopic entropy which deviates from the area law.
\section{State counting for $N=4,8$ black holes}

Let us now finally turn to black hole solutions occuring
in type-IIA compactifications on $K3 \times T^2$ and on $T^6$ and 
let us discuss the associated microstate counting.  In the M-theory
picture we consider then a five-brane wrapped around a
holomorphic four-cycle $P$ in either one of these spaces.
When proceeding with the counting of zero modes, 
as in the Calabi-Yau threefold case described in \cite{MalStrWit}, 
the left- and right-moving bosonic and fermionic degrees of freedom 
are given in terms of the Hodge numbers of $P$ by:
\beqa
N_{\rm bosonic}^{\rm left} &=&  
2 h_{2,0}(P) +  h_{1,1}(P) + 2   - 2 h_{1,0}(P) 
\;,
\nonumber \\
N_{\rm {fermionic}}^{\rm {left}} &=& 4 h_{1,0}(P) \;, \nonumber\\
N_{\rm {bosonic}}^{\rm{right}} &=&  4 h_{2,0}(P) +  4  - 2 h_{1,0}(P) 
\;, \nonumber \\
N_{\rm {fermionic}}^{\rm {right}} &=& 4 [ h_{2,0}(P) + h_{0,0} (P) ] \;.
\label{modes}
\eeqa
The effective two-dimensional theory describing the collective modes of a 
BPS black hole is a $(0,4)$ supersymmetric sigma-model.  
Therefore, the number of right-moving bosons and fermions has to match.
Moreover the right-moving scalars are expected to parametrize a quaternionic
manifold and therefore the number of right-moving real bosons
should be a multiple of four.  Inspection of (\ref{modes})
shows that in the case of a generic Calabi-Yau threefold, for which
$h_{1,0}(P)=0$, 
the counting of right-moving modes is consistent with 
$(0,4)$ supersymmetry, whereas this is not the case
for $K3 \times T^2$ and for $T^6$, for which $h_{1,0}(P)=1$ and
$h_{1,0}(P)=3$, respectively. This implies that the zero-mode counting
for $K3 \times T^2$ and for $T^6$ has to deviate from the one
described above.

Using  (\ref{modes}) the central charges
of the left- and right-moving sector are computed to be 
\beqa
c_{L} &=& N_{\rm {bosonic}}^{\rm{left}} + \ft{1}{2}
N_{\rm {fermionic}}^{\rm {left}} 
=   C_{ABC} \,p^A p^B p^C +
c_{2A} \,p^A + 4 h_{1,0}(P)\,, \nonumber\\
c_{R} &=& N_{\rm {bosonic}}^{\rm{right}} + \ft{1}{2}
N_{\rm {fermionic}}^{\rm {right}} 
=   C_{ABC} \,p^A p^B p^C + \ft{1}{2}
c_{2A} \,p^A + 4 h_{1,0}(P)\,, 
\label{cl}
\eeqa
which then via (\ref{cardmicro}) leads to the following result for the
microscopic entropy,
\beqa
{\cal S}_{\rm {micro}} = 2 \pi \sqrt{ \ft16{|q_{0}|} \Big(
C_{ABC}\, p^{A} p^{B} p^C + c_{2A}\, p^A  + 4 h_{1,0} (P) \Big) } \;.
\eeqa
Now we note that the sub-subleading third term in this expression
proportional to $h_{1,0}$ is not consistent with the macroscopic
computation of the entropy based on $N=2$ supergravity.  As mentioned above, 
the entropy should be even in terms of the charges.

How is the zero-mode counting for $K3 \times T^2$ and for $T^6$ to be modified
in order to remove the inconsistencies mentioned above?  
Let us recall that there are $b_1 = 2 h_{1,0} (P)$ nondynamical gauge fields
present.  If we assume that the zero-modes are charged and couple to 
these gauge fields, then the 
following mechanism suggests itself.  Due to gauge invariance, 
the number of left- and right-moving scalar fields is reduced by $b_1$, 
so that the number of right-moving scalar fields is indeed a multiple of four.
Due to supersymmetry this must be 
accompanied by the removal of $2 b_1$ right-moving
fermionic real degrees of freedom.  If, in addition, we assume that the
removal of fermionic degrees of freedom is left-right symmetric,
then the actual number of left-moving fermionic degrees of freedom is zero.
The central charge in the left-moving sector
is now computed to be $c_L = C_{ABC}\, p^A p^B p^C  + c_{2A}\, p^A$,
which is odd in the charges.
The resulting microscopic entropy formula is then in full agreement
with the macroscopic computation, and it is also 
consistent with anomaly inflow 
arguments \cite{HMM}.



\end{document}